# AIDA: Associative DNN Inference Accelerator

L. Yavits, R. Kaplan and R. Ginosar

**Abstract**— We propose AIDA, an inference engine for accelerating fully-connected (FC) layers of Deep Neural Network (DNN). AIDA is an associative in-memory processor, where the bulk of data never leaves the confines of the memory arrays, and processing is performed in-situ. AIDA's area and energy efficiency strongly benefit from sparsity and lower arithmetic precision. We show that AIDA outperforms the state of art inference accelerator, EIE, by 14.5× (peak performance) and 2.5× (throughput).

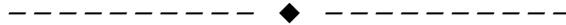

## 1 INTRODUCTION

Fully-connected (FC) layers are extensively used in Convolutional, Recurrent (Long Short-Term Memory) and Multi-Layer Perceptron Neural Networks [1][8]. FC layers perform matrix-vector multiplication (M×V) and non-linear activation function calculation. A number of ASIC and FPGA based platforms that accelerate FC layers exist [1][5][7][8][9][10][11][12].

We propose AIDA, an associative processing in-memory based inference engine. AIDA is a general-purpose massively parallel accelerator that calculates M×V (both sparse and dense matrices and vectors are supported) and non-linear activation function (typically, RELU, sigmoid or tanh), and implements FC of networks such as AlexNet [2] and CTC-3L-421H-UNI [3], outperforming the EIE [1] in terms of peak performance and throughput by 14.5× and 2.5× respectively.

The main sources of AIDA performance and efficiency are (1) ultra-high internal bandwidth achieved by each memory bit being directly connected to processing transistors, (2) associative access that provides intrinsic support for dense and sparse datasets, and (3) associative arithmetic that natively supports variable wordlength and precisions, as well as *perfect induction* approach to implementing arithmetic functions, from multiplication to non-linear activation functions.

This paper makes the following contributions:
- We present the first associative processing in-memory accelerator for FC layers of DNN, that can efficiently support dense and sparse fixed point arithmetic networks;
- We introduce the associative CSR (ACSR) format, a modified CRS format that enables sparsity utilization in associative processor-in-memory.
- We present an algorithm of massively parallel associative in-memory implementation of FC layers.
- We evaluate AIDA on FC layers of CNN and RNN, and compare the performance, throughput and power efficiency of AIDA to those of ASIC and FPGA based accelerators.

The rest of this paper is organized as follows. Section 2 introduces AIDA architecture. Section 3 discusses the FC layer implementation. Section 4 presents the evaluation and Section 5 offers conclusions.

## 2 ACCELERATOR DESIGN

### 2.1 Associative Processing

Associative processor (AP) is a non-von-Neumann in-memory computer [4]. AP is based on Content Addressable Memory (CAM), which allows comparing the entire dataset to a search pattern (key), tagging the matching row, and writing another pattern to all tagged rows. AP performs no computations in conventional sense. Most arithmetic and logic operations can be structured as series of Boolean functions, which are implemented by the AP by perfect induction approach as follows.

The dataset is stored in CAM, typically one data element per CAM row (constituting a Processing Unit, PU). AP controller sequentially matches all possible *input* combinations (hence the name *perfect induction*) of the arguments against the entire CAM content. The matching CAM rows are tagged, and the corresponding function values (precalculated and embedded in AP microcode), are written into the designated *output* fields of the tagged rows.

For an $m$-bit argument $x$ ($x \in$ dataset), any Boolean function $b(x)$ has at most $2^m$ possible values. Therefore, a perfect induction approach would incur up to $O(2^m)$ cycles, regardless of the dataset size.

Arithmetic operations can sometimes (when there are few "don't cares" in the truth table) be more efficiently implemented in a bit-serial, word-parallel manner, reducing time complexity from $O(2^m)$ to $O(m)$ [4].

### 2.2 Hardware Design and Instruction Set

The architecture of AIDA is presented in Fig. 1. The CAM Array comprises bit cells (further described below) organized in bit-columns and word-rows. Several special registers are appended to the CAM array. The COMPARE and WRITE KEY registers contains the patterns (keys) to be compared against and written. The MASK register de-

---
- *Leonid Yavits, E-mail: yavits@technion.ac.il.*
- *Roman Kaplan, E-mail: romankap@gmail.com.*
- *Ran Ginosar, E-mail: ran@ee.technion.ac.il.*

*Authors are with the Department of Electrical Engineering, Technion-Israel Institute of Technology, Haifa 3200000, Israel.*

fines the active fields for write and read operations, enabling bit selectivity. The TAG register marks the rows that are matched by the compare operation and are to be affected by consecutive parallel write. A static associative NOR bitcell is shown in Fig. 1(b). Its two main components are the 6-Transistors (6T) SRAM bit cell (two inverters and 2 write enable transistors) and the 4T wired NMOS XOR.

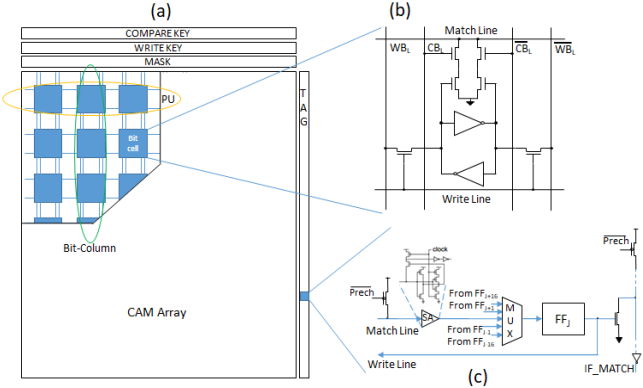

Fig. 1: (a) AIDA Architecture, (b) 10T NOR CAM cell, (c) TAG Logic.

To compare the key data word against the data stored in the CAM array (the entire row, a number of bits or a single bit), the Match line is precharged and the inverted key is set on $CB_L$ and $\neg CB_L$ lines. In the columns that should be ignored during comparison, $CB_L$ and $\neg CB_L$ lines are set to '0'. If all unmasked bits in a row match the key, the Match line remains high and a '1' is sampled into the corresponding TAG FF. If the key differs from the row data (even in one bit), the Match line discharges and a '0' is written into the TAG FF.

In AIDA, compare is typically followed by a parallel write into the unmasked bits of all tagged words. To write data (from the WRITE KEY register) into CAM, each TAG FF (set earlier by the compare) is connected to the corresponding Write line. If a row matched during the compare, the key data set on $WB_L$ and $\neg WB_L$ lines is written into it in accordance with the MASK pattern. Otherwise (in the case of mismatch), the write does not affect the row. AIDA allows simultaneous execution of compare and write operations.

To read data from memory, the $WB_L$ and $\neg WB_L$ lines are precharged and the Write line is asserted.

Two additional AIDA instructions are **if_match**, that signals '1' if there is at least one match in the entire CAM array, and **Move (dir, step)**, that moves tag vector up or down (set by **dir),** by a **step**, which has two values, short_step (1) and long_step (for example, 16, as in Fig. 1(c)).

## 3 ALGORITHM

AIDA implements the following computation:
$$C = f(W \times B) \quad (1)$$
where $W \times B$ is M×V, and $f$ is a non-linear activation function, typically RELU for CNN and sigmoid or tanh for RNN. AIDA implements M×V by inner product method, where all nonzero elements of each row of a weight matrix $W$ are multiplied by nonzero elements of an input activation vector $B$, and dot products are spatially accumulated (reduced) to produce an output activation vector element $C$. The inner product and activation function are calculated in parallel for all rows of weight matrix and all elements of output activation vector, respectively.

The weight matrix $W$ is stored in AIDA's CAM, a single nonzero matrix element per CAM row, in a modified CSR format, which we call *associative CSR* (ACSR) (Fig. 2). The value and the column index vectors are identical to those of conventional CSR, while the row pointer is replaced by a two-bit row flag vector, the elements of which are stored alongside $W$'s values and column indices. Row flag signals the 1st and the last element of a matrix row, as shown in Fig. 2. Every CAM row, a PU, stores nonzero weight matrix, input activation and output activation elements. There is an optional temporary field for storing temporary variables.

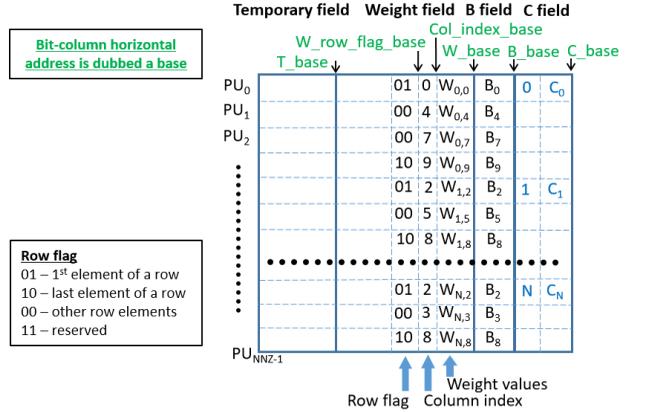

Fig. 2: AIDA CAM map example. The CAM depth is scaled to the weight matrix nnz. Base addresses are detailed in Fig. 3.

The AIDA FC layer algorithm is presented in Fig. 3. It consists of four stages: activation broadcast, multiplication, soft reduction and activation function.

During activation broadcast, each activation either comes from outside (in the case of the first FC layer), or is read from AIDA CAM memory, from the output activation C field (for the subsequent FC layers). It is matched (line 3 of Fig. 3) against and stored in B field (line 4) along all possible multiplication candidates in the weight matrix. Lines 3 and 4 are executed in parallel. Broadcast is done sequentially, activation by activation, but in parallel to all locations of each activation.

In the multiplication step, all nonzero weight-activation pairs are multiplied in parallel. Associative multiplication is a bit-serial operation the dataflow of which is illustrated in Fig. 4(c). The loops in lines 7 and 8 of Fig. 3 go over all bits of the operands. A single-bit multiplication is performed in two steps, a bitwise AND operation and a full addition (lines 9 and 10). A separate bit-column in the temporary field is allocated to store the carry bit during addition. Both bitwise AND and full addition are implemented using the perfect induction approach, by matching each entry of a truth table against the relevant B and W bit-columns, and substituting the result (the value of the function) into the corresponding bits in the temporary field in all matching rows.

**Algorithm:** AIDA FC Layer Implementation

Let W be a weight matrix of *nnz* nonzero elements, stored in ACSR format, as follows:
- $W_{ROW\ FLAG}(nnz)$ is a 2-bit row flag vector, beginning at the row_flag_base;
- $W_{COL\ INDEX}(nnz)$ is a vector of column pointers, beginning at col_index_base;
- $W_{VALUE}(nnz)$ is a vector of nonzero elements of W of wordlength $m$, beginning at W_base;
- Row $j$ of W is held in consecutive CAM rows (PUs), where the 1st PU is marked by $W_{row\ flag}$='01' and the last PU is marked by $W_{row\ flag}$='10'; if there is only one element in row $j$, it is marked by $W_{row\ flag}$='11'

Let B be an input activation vector of $nnz_B$ nonzero elements, presented in the following format:
- $B_{INDEX}$ is a vector of indices;
- $B_{VALUE}$ is a vector of nonzero elements of B of wordlength $n$, beginning at B_base;

Let C be an output activation vector of wordlength $k$, beginning at C_base; Its sign bit resides at C_sign_bit base;
Let Temporary field begin at T_base.

**Main:**
1. **Activation broadcast:**
   // performed in all AIDA PUs in parallel
2.    for (p = 0 ; p < $nnz_B$; p++)  {
3.       Compare ($B_{INDEX}$ ≡ col_index_base);
4.       Write (B_base←$B_{VALUE}$);
5.    }
6. **Multiplication:**
   // performed in all AIDA PUs in parallel
7.    for (j = 0 ; j < n; j++)  {
8.       for (i=0; i < m; i++)  {
9.          Bit_AND(T_base+i+j, W_base+i, B_base+j);
10.         Bit_ADD(C_base+i+j, C_base+i, T_base+j);
11.       }
12.    }
13. **Soft Reduction:**
    // performed in all AIDA PUs in parallel
14.    for (j = 0 ; j < k || j== W_row_flag_base+1; j++) {
15.       Compare (C_base+j ≡ 1);
16.       if ($2^j$<long_step)
17.          Move(up, short_step);
18.       else
19.          Move(up, long_step);
20.       Write (B_base+j ← '1');
21.    }
22.    for (j = 0 ; j < k; j++)  {
23.       Bit_ADD (C_base+j, C_base+j, B_base+j);
24.    }
25.    Compare (W_row_flag≡'10');
26.    if_match goto 'Soft Reduction'
27. **Activation Function (RELU):**
    // performed in all AIDA PUs in parallel
28.    Compare (C_sign_bit≡'1');
29.    Write (C_base← '000...0')

Fig. 3: AIDA FC layer implementation algorithm

If the number of unique multipliers and multiplicands is limited, which frequently occurs in compressed DNN (for example to 16 [1]), the perfect induction method can be applied in a bit-parallel manner, by traversing all possible multiplier-multiplicand combinations and substituting the product values in the matching CAM rows.

The third stage is soft reduction, where the dot products received in the previous stage are accumulated to a final output activation.

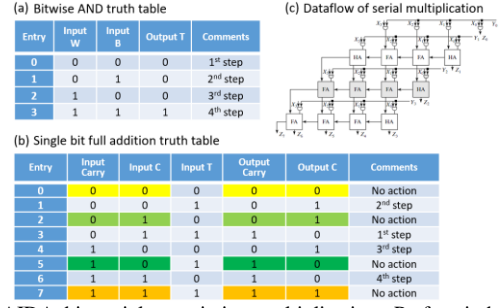

Fig. 4: AIDA bit serial associative multiplication: Perfect induction approach.

Soft reduction is performed in parallel for all output activation elements, where for each such element, we accumulate the dot products in a binary tree fashion: first, all odd and even dot products are summed up; then the odd and even partial results are summed up, and so on, until the last pair of partial results are summed up. Soft reduction requires (1) placing of odd and even dot products and partial results in the same PU, and (2) adding them up. Lines 14 through 20 are responsible for moving the odd dot product or partial result to the PU where the even one is located. Such move is carried out bit-serially, but in parallel in all PUs, in three steps. First, a relevant bit of the source data element is copied to the tag (line 15), then it is shifted to the tag of the destination PU (lines 16 through 19), and finally it is written to the destination bit-column of the destination PU (line 20). The MSB of the corresponding row flag element is shifted along with data. Lines 22 and 23 comprise the addition operation (performed bit-serially in parallel in all PUs). The reduction is performed until all "01" patterns in the row flag vector (signaling the 1st element of each weight matrix row) are replaced by "11" pattern which signals the end of reduction (lines 25 and 26).

In the Activation Function (RELU) stage, we check the sign of all output activations and reset the negative ones (lines 28 and 29). Other activation functions, such as tanh or sigmoid, can be implemented using perfect induction approach.

## 4 EVALUATION

### 4.1 Methodology

We implemented a custom simulator of AIDA, used for performance and power simulation and design space exploration. We compare AIDA with several state of art dedicated hardware (FPGA and ASIC) platforms. A large number of CNN accelerators (DaDianNao [11], Angel-Eye [5], True North [9]), some of which share with AIDA architectural techniques such as bit-serial arithmetic (Cnvlutin [12]), or processing in memory (PRIME [13]), implement both convolutional and FC layers, making comparison difficult. We therefore focus on accelerators where FC layer implementation details are possible to extract. The benchmark networks for AIDA are AlexNet [2] and CTC-3L-421H-UNI [3].

### 4.2 Experimental results

We synthesized the TAG logic (Fig. 1(c)) using Synopsis

Design Compiler with the 45nm FreePDK open cell library and scaled the area and power figures to 28nm technology. A single TAG cell area in 28nm is 7.1μm², and its average energy consumption is 5.6fJ. The 10T NOR CAM bitcell area is 0.135 μm² in 28nm. The estimated area of AIDA capable of storing and processing the compressed FC layers of AlexNet and CTC-3L-421H-UNI is 44.6mm².

TABLE 1. COMPARISON WITH EXISTING PLATFORMS

| Platform | EIE [1] | C-Munk [8] | DNPU [7] | [10] | A-eye [5] | AIDA (ours) |
|---|---|---|---|---|---|---|
| Network | C,R | R | C,R | R | C | C,R |
| Type | ASIC | ASIC | ASIC | FPGA | FPGA | GP/A |
| Memory, B | <8.3M | 82K | 10K | 332K | | 6.4M |
| Freq, MHz | 800 | 168 | 200 | 166 | 150 | 1000 |
| Quant, bit | 4/16 | 8 | 4-7 | 5-16 | 16 | 16/16 |
| Area, mm² | 20.9[1] | 0.27[1] | | | | 44.5 |
| Power, W | 0.37 | 0.029 | 0.02 | 10 | 9.6[2] | 7.15 |
| PP, GOPs | 102 | 32.3 | 25 | 152 | 1.12 | 1474 |
| EE, GOP/J | 2756 | 1114 | 1190 | 15.2 | 14[2] | 206 |
| Thrpt, Inf/s | 81967 | | 1200 | | 33 | 204515 |

*Network, C=CNN, R=RNN; Memory=On chip memory; PP=Peak Performance, EE=Energy Efficiency, Quant=Quantization, Thrpt=Throughput;*
[1] *Area and power are scaled to 28nm*
[2] *Total power (including convolutional and pooling layers)*

Table 1 presents performance, throughput, power and area comparison. All area figures in Table 1 are scaled to 28nm. Large discrepancies in area and power among existing platforms are mainly due to the amount of on-chip memory.

Similar to EIE [1], AIDA utilizes weight and activation sparsity to fit all network parameters in an on-chip memory. AIDA yields the highest throughput, outperforming EIE by 2.5×, however it is 7.7× less energy efficient, because CAM is more power hungry than SRAM.

### 4.3 Design Space Exploration

**Effects of sparsity**: Fig. 5(a) shows AIDA area and energy efficiency as function of weight and activation sparsity, relative to the figures presented in Table 1. Both area and energy efficiency increase almost linearly with weight sparsity.

**Effects of arithmetic precision**: Fig. 5(b) shows AIDA area and energy efficiency as function of weight and activation wordlength, relative to the figures presented in Table 1. Best area and energy efficiency are achieved for binary and ternary networks, and drop with growing weight and activation wordlength. The super-linear growth of energy efficiency is due to quadratic dependency between the associative bit-serial multiplication time and wordlength.

**Scalability**: AIDA operation can be parallelized in a number of ways. For example, the activation broadcast can be performed in parallel with M×V calculations. To accomplish that, the CAM array needs to be partitioned along its rows in two sub-arrays (calculation array and buffer array), each with a separate TAG, connected in every row. This will improve the performance by up to 1.86× while increasing AIDA area by 28%.

Supporting larger networks is fairly straight forward in AIDA. The CAM depth scales to the number of nonzero weights. The CAM width scales according to the weight and activation wordlength. AIDA performance increases sub-linearly to the network size, since the execution time of at least one step of AIDA algorithm, the soft reduction, increases with the number of nonzero output activations, which is likely to grow in larger networks.

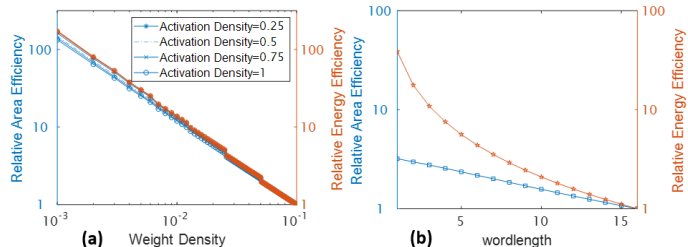

Fig. 5: Effects of sparsity (a) and arithmetic precision (b)

## 5 CONCLUSIONS

This paper presents AIDA, the associative in-memory accelerator of fully connected (FC) layer of neural network. AIDA provides 14.5× higher peak performance and 2.5× higher throughput than EIE [1]. In terms of area and energy efficiency, AIDA strongly benefits from sparsity and lower arithmetic precision. Although in this work we use AIDA as FC layer accelerator, AIDA is in fact a general purpose associative processor in memory, which can efficiently implement a wide variety of data intensive applications, including machine learning, graph processing and sequence alignment.